\documentclass{aa}
\usepackage{graphicx}
\begin{document}
   \title{First XMM-Newton Observations of the Globular Cluster M22}

   \author{Natalie A. Webb 
          \and
           Bruce Gendre
          \and
           Didier Barret 
           }
   \offprints{Natalie.Webb@cesr.fr}

   \institute{Centre d'Etude Spatiale des Rayonnements, 9 Avenue du Colonel Roche, 31028 Toulouse Cedex 04, France }

   \date{Received }

   \abstract{We have examined preliminary data of the globular
   cluster, M22, from the EPIC MOS detectors on board {\it
   XMM-Newton}.  We have detected 27 X-ray sources within the centre
   of the field of view, 24 of which are new detections.  Three
   sources were found within the core of the cluster.  From spectral
   analysis of the X-ray sources, it is possible that the object at
   the centre of the core is a quiescent X-ray transient and those
   lying further out are maybe cataclysmic variables.  }

   \maketitle
         
\section{Introduction}

The nearby (2.6 $\pm$ 0.3 kpc, Peterson \& Cudworth 1994) globular
cluster, M22 (NGC 6656) has previously been studied by the X-ray
satellites {\it Rosat} and {\it Einstein} (see e.g. Verbunt 2001 and
references therein).  Eight X-ray sources (L$_x\ _\sim ^<$ 10$^{34.5}$
erg s$^{-1}$, Hertz \& Grindlay 1983) have already been detected in
the direction of the cluster (Johnston et al 1994), using the {\it
Rosat} PSPC, where the source detected within the core radius is
believed to be related to the cluster.  The eight sources detected by
ROSAT include two of the four X-ray sources detected by {\it Einstein}
(Hertz \& Grindlay 1983).  In this paper we present observations of
M22 taken with the sensitive EPIC MOS camera on board {\it
XMM-Newton}.  We have detected 27 new sources, using this new
instrument, in the direction of the cluster.  Through statistical and
spectral analysis, we have tried to determine the nature of the
sources detected within the core radius and their relationship to the
cluster.

{\bf Globular clusters are known to contain two classes of X-ray
sources: the bright sources which are neutron star low-mass X-ray
binaries and the so-called `dim' sources, with L$_x\ _\sim ^<$
10$^{34.5}$ erg s$^{-1}$ (Hertz \& Grindlay 1983), whose nature is
still discussed.}  The dim sources are intrinsically faint and badly
affected by interstellar absorption, which prevents many sources from
being discovered.  Although the nature of these sources is unclear,
they may be cataclysmic variables (Hertz \& Grindlay 1983) or other
types of binary systems (Verbunt \& Johnston 2000).  It is expected
that globular clusters should contain many binary systems due to the
interactions occurring within the clusters (Di Stefano \& Rappaport
1994).  However only a few such binary systems have been confirmed
(e.g. Verbunt \& Hasinger 1998 and references therein).

\section{Observations and data reduction}

We have obtained approximately 37 kiloseconds (ks) of observations of
the globular cluster M22, with {\it XMM-Newton}.  Observations were
made between 19-20 September 2000, during the `Routine Observing
Phase'.  However, approximately 15 ks were lost due to high background
activity from a solar flare.  We present the data obtained with the
EPIC MOS detectors, using the full frame mode (see Turner et al,
2001).

\begin{figure}
   \includegraphics[angle=0,bb=80 40 690 730,width=14.5cm]{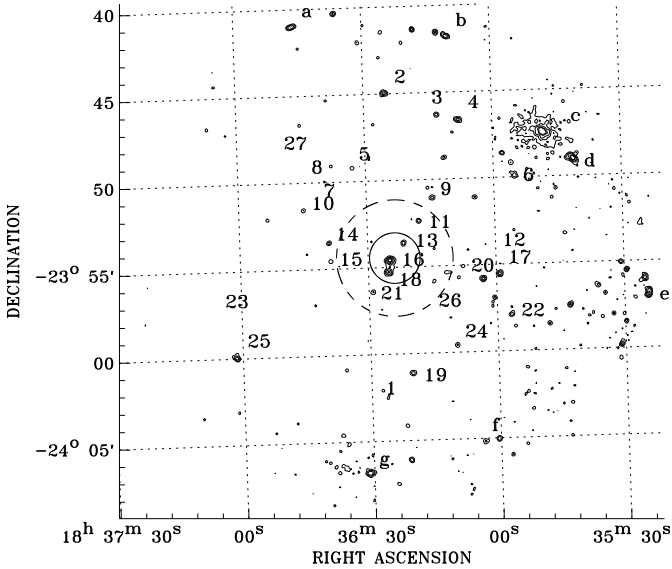}
   \vspace*{-8.0cm}
   \caption{The reduced, 22 ks exposure of M22, in the energy range 0.2-10keV, from the two EPIC MOS cameras. The inner circle (solid line) shows the core radius (85.11'').  The larger circle (dashed line), shows the half mass radius (3.3').  The right ascension and the declination are shown on the abscissa and the ordinate respectively.  The numbers correspond to the ID numbers given in Table~\ref{tab:sources}.  A zoom of the central region can be seen in Figure~\ref{fig:m22zoom}}
   \label{fig:m22mos}
    \end{figure}

The data were processed using the standard MOS pipeline software,
provided in Version 5.0 of the {\it XMM-Newton} SAS (Science Analysis
Software).  Bad pixels were removed from the data, using the task
`badpixfind' and the event lists were filtered, so that 0-12 of the
predefined patterns were retained.  The high background periods were
identified by defining a count rate threshold above the low background
rate and the periods of higher background counts were then eliminated
from the event list.  The resulting background is then very stable
(see also Briel et al, 2001), especially on a small scale, and could
then be subtracted.  The event lists from the two MOS cameras were
merged, to increase the signal-to-noise.  The resulting exposure
corrected image, in the energy range 0.2-10 keV, is presented in
Figure~\ref{fig:m22mos}.

We compare the {\it XMM-Newton} image with the images from the {\it Rosat}
PSPC and HRI observations, obtained from the {\it Rosat} archive at the
Max-Planck-Institut f\"ur extraterrestrische Physik.  The PSPC data,
were taken on 14 March 1991 (8.295 ks) and the two HRI observations
were taken on 16-17 September 1992 and 18-21 March 1993 (9.474 ks
and 31.709 ks respectively).  

\section{The sources}
\label{sec:sources}

The {\it XMM-Newton} sources were first detected using the SAS EPIC
source detection task `eboxdetect', which employs a `local' source
detection algorithm.  A box of 5 $\times$ 5 pixels was used to detect
the point sources and then the same box was used on the background.
To detect the extended sources, two iterations were made with a box of
10 $\times$ 10 pixels and then 20 $\times$ 20 pixels.  A detection
likelihood could then be calculated.

27 sources (see Table~\ref{tab:sources}) have been  detected
using the SAS task `emldetect', using a maximum detection likelihood
of 15. However, only sources in the centre of the field of view (FOV)
have been  detected with this task.  We have  detected a
further 7 sources manually, sources a-g (see Table~\ref{tab:sources}),
at larger off-axis angles, including the extended source, source c
(source A in Hertz \& Grindlay 1983).  These sources all have high
count rates and so we are confident that they are real sources.  
All the detected sources can be seen, numbered, in
Figure~\ref{fig:m22mos}.  In the outer regions of the FOV, some
background fluctuations with multiple contours can be seen, as
fluctuations are magnified in these outer regions, when correcting for
vignetting.  It is for this reason that we have used a maximum
detection likelihood method to detect the sources.

   \begin{figure}
   \includegraphics[angle=0,bb=80 40 690 730,width=14.5cm]{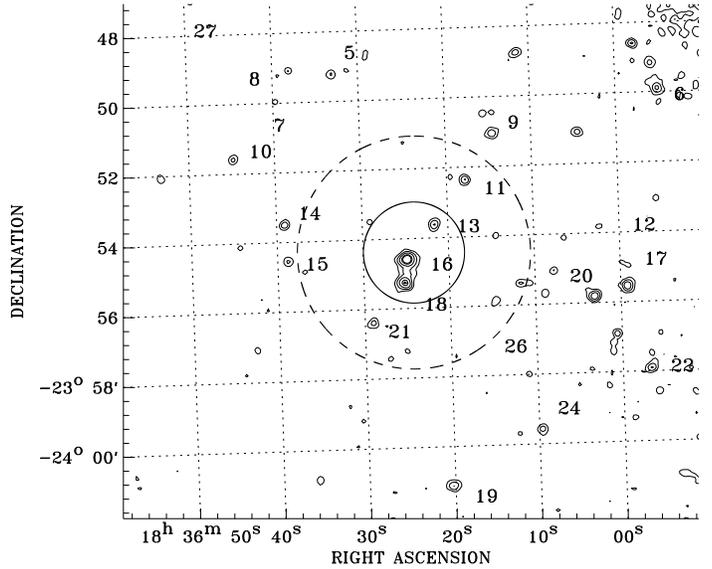}
   \vspace*{-8.0cm}
   \caption{A zoom of the reduced, 22 ks exposure of M22 (Figure~\ref{fig:m22mos}). The inner circle (solid line) again shows the core radius.  The larger circle (dashed line), shows the half mass radius.   The right ascension and the declination are shown on the abscissa and the ordinate respectively.  The numbers correspond to the ID numbers given in Table~\ref{tab:sources}.  The contours describe 3, 6, 11, 20 and 25 sigma certainties.}
   \label{fig:m22zoom}
    \end{figure}

We have detected three sources within the core radius of M22 (85.11'',
Djorgovski 1993), sources 13, 16 and 18 (indicated by a $^*$ in
Table~\ref{tab:sources}), and five sources within the half-mass radius
(3.3', Harris 1996), sources 13, 16, 18, 11 and 21, where previous
observations have detected only one (source 16: using {\it Einstein},
Hertz \& Grindlay 1983; {\it Rosat} PSPC, Johnston et al 1994; and
{\it Rosat} HRI, Verbunt 2001).  The average counts from the MOS 1 and
MOS 2 detectors with their associated errors and the unabsorbed flux
at Earth ($\times$ 10$^{-14}$ ergs cm$^{-2}$ s$^{-1}$) in the energy
band 0.5-2.5 keV are given in Table~\ref{tab:sources}.  This energy
band has been chosen, so that the results are comparable with former
source detections, e.g. Johnston et al (1994).  We have assumed a 5
keV exponential spectrum, using the column density $N_H = 2.2 \times
10^{21} {\rm cm}^{-2}$, and a distance of 3 kpc, in the same way as
Johnston et al (1994).  The approximate position of each source
detected and any former detection is also given.

\begin{table}
  \caption[]{X-ray sources in the direction of M22, as determined from the EPIC MOS observations}
     \label{tab:sources}
     $$ 
       \begin{array}{lccccc}
         \hline
         \noalign{\smallskip}
           & For. & RA (2000) & Dec (2000) & Average & Flux  \\  
           ID & ID & ^h \hspace*{3mm} ^m \hspace*{3mm} ^s & ^{\circ} \hspace*{3mm} ' \hspace*{3mm} '' & Counts & \times10^{-14} \\
         \hline
         1 &  & \ 18\ 36\ 24.0 & -24\ 02\ 21.0 & 25\pm6 & 1.04 \\
         2 &  & \ 18\ 36\ 27.9 & -23\ 44\ 58.4 & 47\pm7 & 1.95 \\
         3 &  & \ 18\ 36\ 15.4 & -23\ 46\ 06.7 & 26\pm6 & 1.08 \\
         4 & 1 & \ 18\ 36\ 10.5 & -23\ 46\ 19.6 & 52\pm8 & 2.16\\
         5 &  & \ 18\ 36\ 34.8 & -23\ 49\ 19.2 & 28\pm6 & 1.16 \\
         6 &  & \ 18\ 35\ 56.8 &  -23\ 49\ 20.6 & 24\pm6 & 1.00 \\
         7 &  & \ 18\ 36\ 40.0 & -23\ 50\ 48.9 & 23\pm5 & 0.95 \\
         8 &  & \ 18\ 36\ 05.9 & -23\ 50\ 43.1 & 27\pm6 & 1.12 \\
         9 &  & \ 18\ 36\ 15.8 & -23\ 50\ 51.1 &29\pm6 & 1.20 \\
         10 &  & \ 18\ 36\ 45.7 & -23\ 51\ 48.0 & 29\pm6 & 1.20 \\
         11 &  & \ 18\ 36\ 18.6 & -23\ 52\ 09.7 & 29\pm6 & 1.16 \\
         12 &  & \ 18\ 36\ 02.7 & -23\ 53\ 21.2 & 23\pm5 & 0.95 \\
         13^*  &  & \ 18\ 36\ 21.9 & -23\ 53\ 30.7 & 40\pm7 & 1.66 \\
         14 &  & \ 18\ 36\ 39.3 & -23\ 53\ 40.9 & 43\pm7 & 1.78\\
         15 &  & \ 18\ 36\ 38.7 & -23\ 54\ 39.5  & 26\pm6 & 1.08 \\
         16^* & B,4 & \ 18\ 36\ 25.0 & -23\ 54\ 32.3 & 193\pm13& 8.01 \\
         17 &  & \ 18\ 35\ 59.1 & -23\ 54\ 58.4 & 45\pm7 & 1.87 \\
         18^*  &  & \ 18\ 36\ 25.2 & -23\ 55\ 08.8 & 92\pm9 & 3.82  \\
         19 &  & \ 18\ 36\ 18.5 & -24\ 00\ 52.8 & 42\pm7 & 1.74 \\
         20 &  & \ 18\ 36\ 03.0 & -23\ 55\ 21.4 & 55\pm7 & 2.28 \\
         21 &  & \ 18\ 36\ 28.6 & -23\ 56\ 22.8  & 23\pm6 & 0.95\\
         22 &  & \ 18\ 35\ 55.6 & -23\ 57\ 14.1 & 20\pm5 & 0.83\\
         23 &  & \ 18\ 37\ 00.8 & -23\ 58\ 24.4  & 22\pm5 & 0.91 \\
         24  & & \ 18\ 36\ 08.5 & -23\ 59\ 13.9 & 21\pm5 & 0.87 \\
         25  & & \ 18\ 37\ 00.4 & -24\ 00\ 22.9  & 56\pm7 & 2.32 \\
         26  & & \ 18\ 36\ 19.3 & -23\ 57\ 10.0 & 26\pm6 & 1.08 \\
         27 & 9 & \ 18\ 36\ 49.0 & -23\ 48\ 02.7 & 42\pm9 & 1.74 \\
         \noalign{\smallskip}
         \hline
         \noalign{\smallskip}
         a  &  & \ 18\ 36\ 50.5 & -23\ 41\ 24.5 & 47\pm9 & 1.95 \\
         b  &  & \ 18\ 36\ 14.1 & -23\ 41\ 38.6 & 64\pm10 & 2.66 \\
         c & A,3 & \ 18\ 35\ 50.6 & -23\ 46\ 49.4 & 256\pm17 & 10.62 \\
         d  &  & \ 18\ 35\ 43.0 & -23\ 48\ 18.3 & 145\pm13 & 6.02 \\
         e  & 5 & \ 18\ 35\ 23.8 & -23\ 55\ 52.0 & 171\pm16 & 7.10 \\
         f  & 7 & \ 18\ 36\ 00.5 & -24\ 04\ 41.1 & 44\pm8 & 1.83 \\  
         g  & 8 & \ 18\ 36\ 27.3 & -24\ 06\ 48.2 & 72\pm9 &  3.00  \\  
         \noalign{\smallskip}
         \hline
      \end{array}
     $$ 
    \begin{list}{}{}
      \item ID: 1-27, sources in the centre of the EPIC MOS FOV.
      \item \hspace*{0.5cm}  a-g at the edge of the EPIC MOS FOV.
      \item For. ID: A-B, {\it Einstein}, Hertz \& Grindlay 1983
      \item \hspace*{1.3cm} 1-8, {\it Rosat} PSPC, Johnston et al 1994
      \item \hspace*{1.3cm} 9, {\it Rosat} HRI only, Verbunt 2001
      \item $^*$ denotes sources within the core
    \end{list}
  \end{table}

The dataset received was incomplete, so the astrometry was improved
using the 3 sources that were detected both in our data and by the
{\it Rosat} HRI (sources 16, 27 and e, Table~\ref{tab:sources}).  An
IDL routine was used to derive the adjustment required for the MOS
data to align with the HRI data.  We found that we required both a
small transversal shift and rotation, in the same manner as Hasinger
et al (2001).  This leads to a residual error of $\sim$5.5'', where the
largest error is due to the position error of the sources detected by
the HRI.

The six brightest sources detected by the {\it Rosat} PSPC have also
been detected by the EPIC MOS detectors.  However, the two weakest
sources detected by the PSPC are not apparent in our data.  We have
searched to a limiting luminosity of 3.0 $\times$ 10$^{30}$ ergs
s$^{-1}$ in the 0.5-2.5 keV band, derived in a similar manner as
Johnston \& Verbunt (1996), using the upper limits of 3 positions, in
or near to the core.  As the {\it Rosat} PSPC has a lower limiting
luminosity of 1.3 $\times$ 10$^{31}$ ergs s$^{-1}$ in the 0.5-2.5 keV
band (Johnston \& Verbunt 1996), but the maximum likelihood of
existence for each is $_\sim ^>$10 (Johnston et al, 1994), the two
faintest sources detected by the PSPC are likely to be variable
sources.

\section{Cluster Membership}
\label{sec:member}

It is unclear how many of the sources that we have detected are indeed
members of the cluster.  Searching the SIMBAD database, none of the
X-ray sources have an optical counterpart that lies within the error
box of the X-ray determined coordinates.  We have therefore tried to
evaluate whether we see an overdensity of sources within the core of
the cluster, in a similar manner to Fox et al (1996).  We have used
the log N-log S relation derived from {\it XMM-Newton} observations of
the Lockman Hole (Hasinger et al 2001).  Our lowest detected source
has an unabsorbed flux value of 8.3 $\times 10^{-15}$ ergs cm$^{-2}$
s$^{-1}$ (0.5-2.5 keV).  In the same energy range (0.5-2.0 keV) as the
log N-log S relationship, this is 6.0 $\times 10^{-15}$ ergs cm$^{-2}$
s$^{-1}$.  Using a spectrum with a photon index of 2.0, the photon
index given for sources in the energy band 0.5-2.0 keV (Hasinger et al
2001), rather than the 5 keV exponential spectrum that we have used to
determine the flux, the flux drops to 3.3 $\times 10^{-15}$ ergs
cm$^{-2}$ s$^{-1}$.  This indicates that we should see 220 sources per
square degree, using the log N-log S relationship.  However the area
of the core is 1.76 $\times 10^{-3}$ square degrees, thus we expect
0.4 sources within the core.   According to Poisson statistics,
using the expected number of sources and one trial, the probability of
a chance detection of one source in the core is $\sim$27\%.

In addition, it is also possible to calculate the probability that the
sources within the core are not simply spurious identifications with
the cluster, in the same manner as Verbunt (2001).  Using the
probability, $p$, that one serendipitous source in the MOS observation
is at a distance $r < R$ from the cluster centre, located at RA=18$^h$
36$^m$ 24.2$^s$, dec=-23$^{\circ}$ 54' 12'' for M22 (Djorgovski \&
Meylan 1993), where $p = (R/r_d)^2$ and $r_d$ is the radius of the
field of view.  For our observations with {\it XMM-Newton}, we
consider only the central field of view (radius=0.134${^\circ}$),
where 27 sources have been automatically detected (see
Section~\ref{sec:sources}). The probability of  not finding any
sources  in a single trial simply by chance, within the core
radius, is 97\%.   In 27 trials, for the 27 sources, the
probability of finding no sources within the core radius is 44\%.  It
is therefore possible that the faintest source within the core radius
(source 13) may not be related to the cluster, where from their
brightness, the other two core sources are more likely to be related. 

We also consider the sources within the half-mass radius, as these
sources are likely to be associated with the cluster. Both the {\it
Einstein} observations, L$_{min}$=1.02 $\times$ 10$^{32}$ ergs
s$^{-1}$ in the 0.5-2.5 keV band (source D), with a limiting
luminosity of the observations of approximately 1.0 $\times$ 10$^{32}$
ergs s$^{-1}$ (Hertz \& Grindlay 1983) and the {\it Rosat} PSPC data,
where the limiting luminosity is 1.3 $\times$ 10$^{31}$ ergs s$^{-1}$
in the 0.5-2.5 keV band (Johnston et al 1994) have detected only 1
source within the half-mass radius.  Scaling the lowest observed
luminosity in a similar way as we have already done, and using the
same log N-log S relation derived from {\it XMM-Newton} observations
of the Lockman Hole (Hasinger et al 2001), for the half-mass radius of
the {\it Einstein} observations, we should see approximately zero
X-ray sources in such an area and similarly for the {\it Rosat}
observations.  For our observations with {\it XMM-Newton}, 2 sources
would be expected, thus we see an overdensity of 3 X-ray sources
within the half-mass radius.  {\bf If the sources detected within the
half-mass radius are at the distance of the cluster, their
luminosities range from 1.0-8.7$\times$ 10$^{31}$ ergs s$^{-1}$, in
the 0.5-2.5 keV band, where 8.7$\times$ 10$^{31}$ ergs s$^{-1}$ is the
luminosity of the most central source, source 16.}

\section{Variability}

X-ray variability can occur on different timescales, depending on the
source of the variability, variations from milliseconds (e.g. from
the neutron star spin and orbital motion close around the neutron
star, van der Klis 2000) to hours (e.g. the spin period of the white
dwarf in OY Car (Ramsay et al 2001).  It can therefore be a useful
diagnostic of the nature of the object observed.

We have attempted a timing study of source 16, which has the highest
count rate of all the objects in the centre of the field of view.
However, using the counts from all the energy bands, over the 37 ks,
results in only 0.011 counts per second.  We have binned the counts
into bins of several hundred seconds and performed a period search
using the time-series analysis package `period'.  However, we found no
significant period in the data, other than the period on which the
data were binned, and aliasses thereof.  

Binning the longest continuous sample of data into 4 data bins, where
each bin contains approximately 85 counts, no variability can be seen,
larger than the size of the error bars.  We therefore require
further counts to carry out a proper study.

\section{Spectra}

   \begin{figure}
   \includegraphics[angle=-90,bb=90 140 600 650,width=9cm]{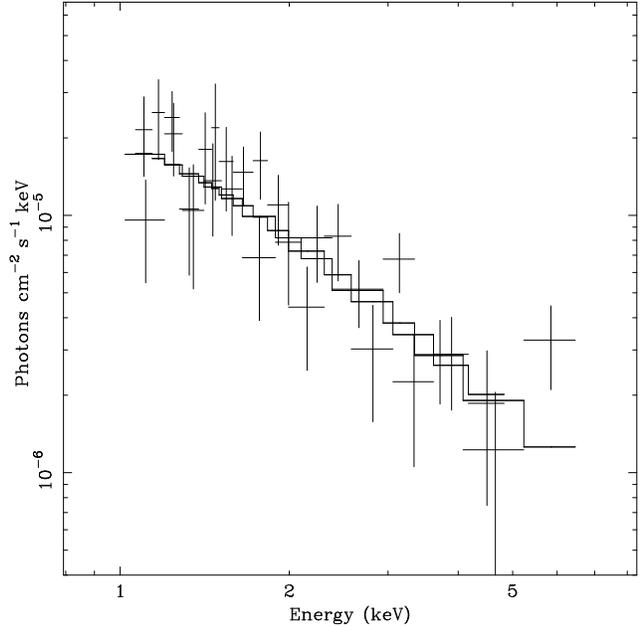}
   \caption{Spectrum of source 16, between 1-7 keV. Data from both
   the EPIC MOS cameras are plotted, where each data bin contains at
   least 20 counts.  The solid lines show the best absorbed power law
   fit.}  
  \label{fig:spec} 
\end{figure}

We have fitted the spectra of the sources within the cluster core,
between 0.2-10 keV.  Two of these sources are two of the brightest of
the 27 sources  detected in the centre of the field of view.  For all
of the fits, we have used a column density of $N_H=2.16 \times 10^{21}
{\rm cm}^{-2}$, as was used in Johnston et al (1994).

The best fit absorbed power law spectrum to the most central source,
source 16 (Table~\ref{tab:sources}) is shown in Figure~\ref{fig:spec}.
As the channel boundaries are different for MOS1 and MOS2, the fit is
duplicated, as can be seen in Figure~\ref{fig:spec}, although the fit
is indeed the same in both cases.  The data has been binned up, to
contain at least 20 counts in each bin.  The fit with the lowest
$\chi^{\scriptscriptstyle 2}_{\scriptscriptstyle \nu}$ for the above
data is a power law with a spectral index of 1.84$\pm$0.17
($\chi^{\scriptscriptstyle 2}_{\scriptscriptstyle \nu}$=0.983, 27
degrees of freedom). All the errors quoted are at the 90\% confidence
limit.  However we obtain almost an equally good fit for a single
component, bremsstrahlung model, with the parameters kT=4.51$\pm$1.51
keV ($\chi^{\scriptscriptstyle 2}_{\scriptscriptstyle \nu}$=1.004).
The poorest, yet still acceptable fit is a blackbody, the source has a
temperature of 0.58$\pm$0.04 keV ($\chi^{\scriptscriptstyle
2}_{\scriptscriptstyle \nu}$=1.406). We calculated the hardness ratios
for the regions 3.0-10.0/1.0-3.0 keV and 1.0-3.0/0.5-1.0 keV.  These
show a hard spectrum with no high energy cutoff, with values of
0.52$\pm$0.09 and 3.05$\pm$0.99 for the two ratios respectively.

Sources expected are soft X-ray transients in quiescence (Verbunt
\& Hut 1987), cataclysmic variables, RS CVn binaries, and recycled
neutron stars (Verbunt \& Johnston 2000).  Bahcall \& Lightman (1976)
(see also Lightman et al 1980; Grindlay et al 1984) discuss the fact
that an X-ray source in a globular cluster can be treated as a test
particle in thermal equilibrium with the other cluster stars.  Thus,
the probability distribution for the location of the X-ray source
depends on its mass, with the more massive objects lying nearer to the
centre of the cluster.

The most massive object expected in a globular cluster, is a quiescent
X-ray transient, containing a black hole.  In globular clusters with a
high density of stars, the dynamical evolution is so rapid that 
many black holes will be ejected from the system on relatively short
time-scales (Portegies Zwart \& McMillan 2000; Kulkarni et al
1993),  although some of these will be retained in the cluster
and may capture a `normal star' to form a low mass X-ray binary, with
occasional X-ray outbursts (Kulkarni et al 1993).  

This spectrum is similar to the black hole X-ray transient spectra
depicted in Asai et al (1996), Asai et al (1998) and Campana \& Stella
(2000). If the most central object were a black hole X-ray transient,
it would be likely that the compact object accretes via an advection
dominated accretion flow (ADAF), see Narayan et al (1996), where it is
proposed that such objects would show hard X-ray spectra.

Neutron stars have a mass of approximately 1.4M$_\odot$ (e.g. Thorsett
et al, 1993), and thus an X-ray binary in such a globular cluster
would have a lower mass if the black hole were replaced with a neutron
star.  The observed spectrum has a blackbody temperature higher than
that of the proposed identification of a quiescent transient neutron
star in the globular cluster NGC 5139, and other similar objects
examined (Rutledge et al 2001), although the hard spectral tail has a
power law photon index that is consistent with quiescent transient
neutron stars known (Rutledge et al 2001 and Asai et al 1996).
Alternatively, globular clusters are already known to be a rich source
of millisecond pulsars (D'Amico et al 2001).  12 millisecond pulsars
(MSP) have also recently been detected by the X-ray telescope {\it
Chandra} in the globular cluster 47 Tuc (Grindlay et al 2000).  Our
spectral analysis does not exclude the possibility that source 16 is a
MSP.

X-ray transients have a higher minimum mass than cataclysmic
variables, where the minimum mass of a known cataclysmic variable (CV)
is approximately 0.5M$_\odot$ (Ritter \& Kolb 1998).  A large number
of CVs are expected to form in globular clusters through tidal capture
(Hertz \& Wood 1985) and therefore to exist and contribute to the
number of low-luminosity X-ray sources observed in such clusters (Di
Stefano \& Rappaport 1994).  RS CVn binaries in a globular cluster
contain a late-type, usually evolved, (spectral type $\sim$K/M) star.
For M22, where the turn-off mass is about 0.8M$_\odot$ (Piotto \&
Zoccali 1999), RS CVn binaries within the cluster should have a
similar mass to CVs in the cluster.

We have detected a second source (source 18) at the centre of the
globular cluster, only 36'' from source 16 (source B in Hertz \&
Grindlay 1983).  Source 16, in the core of the cluster, has been
detected by both {\it Einstein} and {\it Rosat}, but in both cases, as
a single source.  This second source is, however, too faint to have
been detected by the deepest survey, by the {\it Rosat} PSPC, Johnston
et al (1994).  Fitting this source in a similar manner, we find that
a power law fit with a spectral index of 1.36$\pm$0.26
($\chi^{\scriptscriptstyle 2}_{\scriptscriptstyle \nu}$=1.357, 12
degrees of freedom) gives the best $\chi^{\scriptscriptstyle
2}_{\scriptscriptstyle \nu}$.  A blackbody with a temperature of
0.76$\pm$0.09 keV ($\chi^{\scriptscriptstyle 2}_{\scriptscriptstyle
\nu}$=1.875), can also be used to describe the data. However, we find
that a bremsstrahlung model is not a good description of the data.  We
calculated the hardness ratios in the same way as for source 16.  The
hardness ratios, 0.29$\pm$0.07 and 3.36$\pm$1.38 respectively, show
that the two sources, 16 and 18, are indeed different.

As a large number of cataclysmic variables are expected in globular
cluster systems, it could be expected that this source is a
cataclysmic variable.  However, the spectrum is not typical of a CV,
e.g. Richman 1996, who finds, in general, that a bremmstrahlung model
with a temperature of kT=0.1-5 keV is a good description of the X-ray
spectra of many CVs.  However, Richman also finds, in the case of case
of GQ Mus, that a low temperature blackbody is the best description.
Van Teesling \& Verbunt (1994) find that a wider range of models fit
the CV spectra, including multiple component fits.

The spectra of sources 16 and 18, are significantly different,
especially the hardness ratios, supporting the fact that they are
indeed different, individual sources.  The third source in the cluster
core has too few counts to be fitted accurately.

\section{Conclusion}

We have examined preliminary data from the EPIC MOS detectors on board
{\it XMM-Newton} of the globular cluster M22.  We have  detected 27
X-ray sources within the field of view, in the line of
sight to M22 and a further 7 sources at larger off-axis angles.  These
indentifications include 7 of the sources formerly detected, but we
find no evidence for the two faintest sources found by the {\it Rosat}
PSPC camera.  We have detected three sources within the core of M22,
where previous observations have  detected only one.

It is unclear which sources are associated with the cluster, as none
of the sources coincide with known optically identified sources,
although it is likely that all 3 of the sources detected within the
core radius are associated with the cluster.  We have presented
evidence, through the X-ray spectra, to indicate that the X-ray source
at the centre of the core may be a quiescent X-ray transient and that
the two sources lying further out may be cataclysmic variables.

\begin{acknowledgements}
We are grateful to M. Auri\`{e}re and J. Ballet for their comments
during the preparation of this manuscript.   We are also very grateful
to the referee, Frank Verbunt, for many helpful comments.

\end{acknowledgements}

\end{document}